\documentclass[12pt]{article}
\usepackage[utf8]{inputenc}
\usepackage{amsmath}
\usepackage{amsfonts}
\usepackage{amssymb}
\usepackage{epsfig}
\usepackage{epstopdf}
\usepackage{indentfirst}
\usepackage{geometry}
\usepackage{graphicx}
\usepackage{caption}
\usepackage{subfig}
\usepackage[countmax]{subfloat}
\usepackage{rotating}
\usepackage{lscape}
\usepackage[colorlinks]{hyperref}
\usepackage{tabularx}
\usepackage{lineno}

\newtheorem{theorem}{Theorem}[section]
\newtheorem{lemma}{Lemma}[section]

\newtheorem{proposition}{Proposition}[section]

\newtheorem{definition}{Definition}[section]

\numberwithin{equation}{section}
\def\bbR{{\mathbb R}}

\def\cF{{\mathcal F}}
\def\cL{{\mathcal L}}

\graphicspath{%
    {converted_graphics/}
    {/}
}

\begin{document}

\title{How dark is the dark side of diversification?}
\author{Pedro Cadenas$^1$, Henryk Gzyl$^2$ and Hyun Woong Park$^1$\\
$^1$ Department of Economics, Denison University, Granville.\\
$^2$ Center for Finance, IESA, Caracas.\\
{\small cadenasp@denison.edu; henryk.gzyl@iesa.edu.ve; parkhw@denison.edu}
} 

\date{}
 \maketitle

\baselineskip=1.5 \baselineskip \setlength{\textwidth}{6in}

\begin{abstract}
Against the widely held belief that diversification at banking institutions contributes to the stability of the financial system, Wagner (2010) found that diversification actually makes systemic crisis more likely. While it is true, as Wagner asserts, that the probability of joint default of the diversified portfolios is larger; we contend that, as common practice, the effect of diversification is examined with respect to a risk measure like VaR. We find that when banks use VaR, diversification does reduce individual and systemic risk. This, in turn, generates a different set of incentives for banks and regulators. In other words, the decisions made by banks and regulators with regards to diversification depends on how risk is assessed. We explore some of the implications of these results for the financial system. Along the way we extend Wagner's results of his model, and make a probabilistic analysis that relates the probability of joint default to the correlations between the diversified portfolios. 
\end{abstract}
\par\vspace{6pt}

\par\vspace{6pt}

\section{Introduction}

Wagner's (2010) linear model shows that although diversification reduces the individual probability of failure of each banking institution, it comes at the cost of making systemic crisis more likely.  In generic terms, Wagner found that if one considers two portfolios made up of two independent assets, the joint probability of the portfolios being less than a certain amount (call it $d$), is larger than the joint probability of the individual assets being less than $d.$ 

\par\vspace{6pt}

Wagner (2010) considers a system consisting of two banks, whose assets are described by two independent random variables $X$ and $Y,$ uniformly distributed in $[0,s].$ The model then examines what happens when the two banks rearrange their assets to be $\nu(1) = (1-r_1)X + r_1Y$ and $\nu(2) = r_2X + (1-r_2)Y.$ Then he computes $P(\nu_1 \leq d,\nu_2\leq d)$ and proves that it is larger than $P(X \leq d,Y\leq d),$ where $d$ is some threshold below which any bank is said to enter default. On the basis of this comparison, Wagner concludes that diversification comes at a cost. As a consequence, Wagner (2010) shows that there is a rationale for discouraging diversification. 

\par\vspace{6pt}

We argue that Wagner's model may also lead to a different rationale if banks and regulators use a risk measure like Value at Risk (VaR). In particular, we show that since the probability of joint default is not a risk measure, diversification may not necessarily be perceived as increasing the risk of failure of the banking system. Hence, if VaR is used by banks and regulators the rationale for discouraging diversification disappears. There is nothing wrong with Wagner's analysis. As a mater of fact we extend his arguments and examine the dependence of the probabilities on the joint pair $(r_1,r_2);$ and contribute to further refine the analysis by offering a bridge between the model and the covariance $Cov(\nu_1,\nu_2)$ of the banks portfolios. This connection is quite important in Wagner's model because the likelihood of systemic crisis increases as banks become more similar to one another. The analysis of VaR in the context of Wagner's model (2010) that we present here, serves to illustrates the point that the way in which we assess the benefit of diversification, and its dark side (costs), ultimately depends on the risk metric that is being adopted.



\par\vspace{6pt}

The paper is organized as follows, in Section 2 we examine the probability of default of an investor with a diversified portfolio. We shall see that for appropriate diversification, the probability of default can be smaller than the probability of one of the banks defaulting with no diversification. When there are two investors, the probability of a joint default is larger than the probability of joint default of the two banks, but this might be traced to the fact that the returns of the two diversified investors are not independent any more. We examine how does the covariance $Cov(\nu_1,\nu_2)$ depend on $(r_1,r_2);$ and on the identical and independent distribution of the asset returns assumption used in Wagner (2010). In Section 3 we prove that for appropriate redefined positions, the value at risk of a diversified portfolio is smaller than that of the individual risks. We also show that if banks use the same degree of diversification, the value at risk of the entire banking system is less than the VaR of the entire banking system without diversification. In section 4 we explore the implications for banks and regulators who discourage or encourage diversification by changing the level of capital requirements.

\section{The model}
\begin{figure}[h!]
    \centering
    \subfloat[$r_i< \frac{d}{s}$]{
		\includegraphics[width=0.4\textwidth]{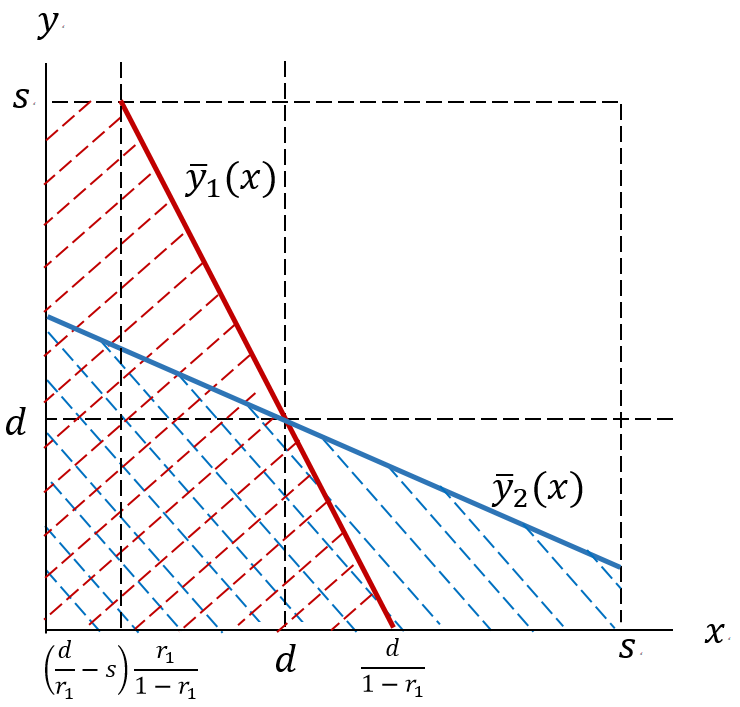}
		\label{fig:wagner1}}
    \subfloat[$r_i \geq \frac{d}{s}$]{
		\includegraphics[width=0.4\textwidth]{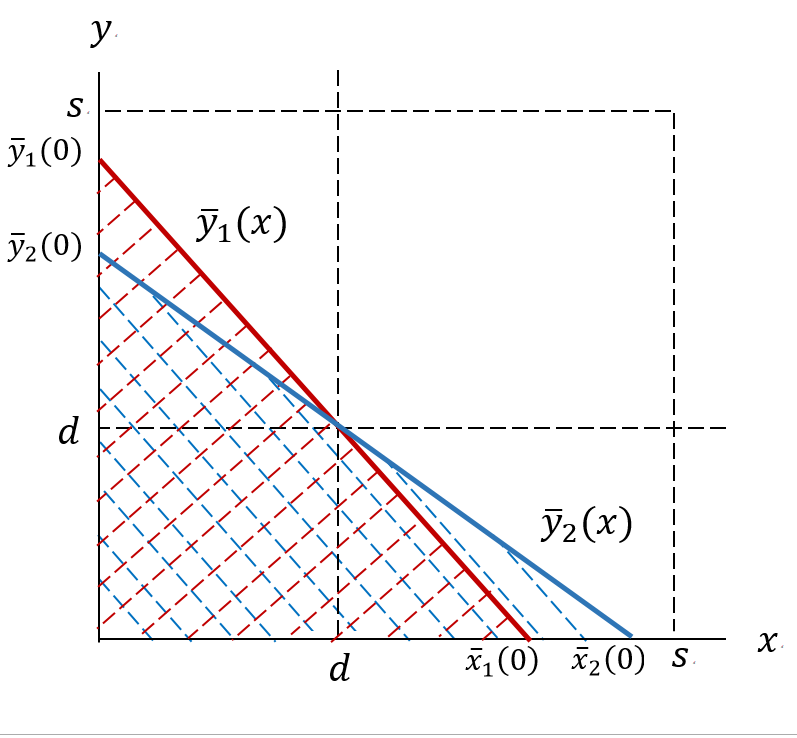}
				\label{fig:wagner2}}
\caption{The bank run outcomes}
\label{fig:wagner}
\end{figure}

Using the same notations as in \cite{W} consider the positions of the two investors with diversified portfolios given by:
$$\nu(1) = (1-r_1)X + r_1Y$$
$$\nu(2) = r_2X + (1-r_2)Y$$
We shall say that they loose money whenever $\{\nu(1)<d\}$ or $\{\nu(2)<d\},$ which are the regions in the $(X,Y)$ plane (respectively) bounded by the lines $(1-r_1)x + r_1y=d$ and $r_2x + (1-r_2)y=d$. Two possible configurations of these lines emerge as shown in Figure \ref{fig:wagner}. It is easy to verify that the dividing line occurs when $y_1 (0)=s$ for bank 1 and $x_2 (0)=s$ for bank 2, both of which correspond to $r_i = \frac{d}{s}$. We consider each of these in our analysis below.

\subsection{The probability of individual failure}
The first case of bank run outcomes, displayed in Figure \ref{fig:wagner}\subref{fig:wagner1}, is when $r_i < \frac{d}{s}$. In this case, the default probability of, say, bank 1 is the integral over the red area in figure. Under the uniform distribution assumption, with density $1/s$, we have 
\[P(\nu_1 <d) = \left[\left(\frac{d}{r_1}-s\right)\frac{r_1}{1-r_1}s+ \left\{\frac{d}{1-r_1}- \left(\frac{d}{r_1}-s\right)\frac{r_1}{1-r_1}\right\}\frac{s}{2}\right]\frac{1}{s^2}\]
which results in, 
\begin{equation}
P(\nu_1 <d) =\frac{2d -r_1 s}{2s(1-r_1)}
\end{equation} 
Confirm that $r_1 = 0$ (no diversification) makes $P(\nu_1 <d) = \frac{d}{s}$. Same for bank 2's case. Hence, we have
\begin{equation}
P(\nu_i <d) = \frac{2d -r_i s}{2s(1-r_i)}
\end{equation}
The probability of individual failure when banks diversify is less or equal than the probability of individual failure when the banks do not diversify, if the following condition holds:
\begin{equation}
P(\nu_i<d) = \frac{2d -r_i s}{2s(1-r_i)} \leq \frac{d}{s}
\end{equation}
which simplifies to
\begin{equation}
s \geq 2d. \label{s>2d}
\end{equation}
In all, as long as the condition \eqref{s>2d} holds, any $r_i \in (0,d/s)$ will make a bank failure less likely compared to when the bank does not diversify.
 
 \par\vspace{6pt}
 
The second case of bank run outcomes, displayed in Figure \ref{fig:wagner}\subref{fig:wagner2}, is when $r_i \geqslant \frac{d}{s}$. The boundary lines cut the axes at:
$$(x_1(0),y_1(0)) = \left(\frac{d}{1-r_1},\frac{d}{r_1}\right)\;\;\;(x_2(0),y_2(0)) = \left(\frac{d}{r_2},\frac{d}{1-r_2}\right)$$

As before, the default probability of bank 1 is the integral over the red area in figure and similarly for bank 2's case. Accordingly, under the uniform distribution assumption, we get
\begin{equation}
P(\nu_i<d) = \frac{1}{2}\Big(\frac{d}{s}\Big)^2\frac{1}{r_i(1-r_i)} \label{cc2}
\end{equation}		
For diversification to reduce, or make it equal to, the probability of individual failure when compared with the no diversification strategy---formally, $P(\nu_1<d) \leq P(X <d)$ and $P(\nu_2<d) \leq P(Y < d)$)---$r_i$ should be such that,
\begin{equation}
P(\nu_i) = \frac{1}{2}\left(\frac{d}{s}\right)^2\frac{1}{r_i(1-r_i)} \leq \frac{d}{s}
\end{equation}
Or equivalently, 
\begin{equation}
\frac{d}{2s} \leq r_i(1-r_i) \label{eq:4}
\end{equation}

\begin{figure}[h!] 
  \centering
  \includegraphics[width=3.0in,height=2.in]{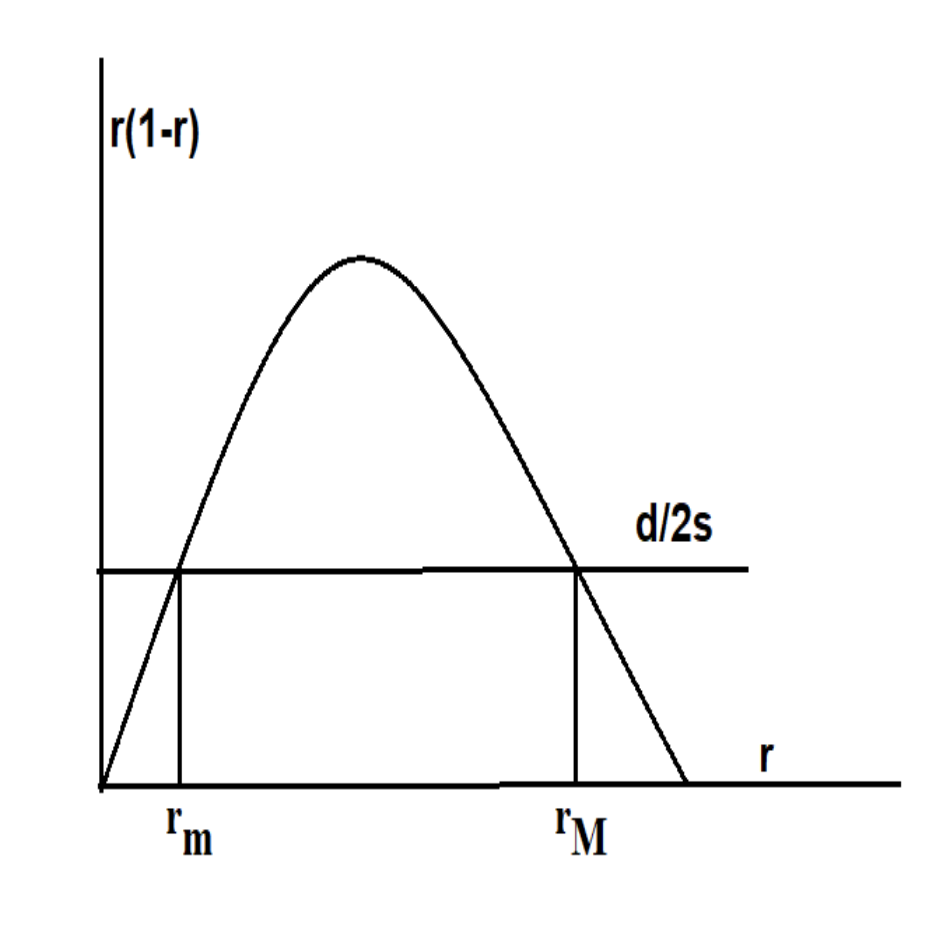}
  \caption{Region of no default for individual investor ($r_i \geq \frac{d}{s}$)}
  \label{Prob}
\end{figure}

The relation is visualized in Figure \ref{Prob}, from which we can tell that the relevant degree of diversification is $r_i \in [r_m, r_M]$. Since $r_i(1-r_i) \leq 1/4$, the condition \eqref{eq:4} implies 
\begin{equation}
s \geq 2d
\end{equation}

Solving the above quadratic equation yields $r_m, r_M = \frac{1}{2}\left[1\pm \sqrt{1-\frac{2d}{s}} \right]$. Due to $s \geq 2d$, we have $d/s < r_m 	\leq 1/2$ and $1/2 \leq  r_M \leq 1$. 

\par\vspace{6pt}

In all, the following proposition summarizes the discussion thus far.
\begin{proposition}
Bank $i$'s degree of diversification that ensures $P (v_i <d) \leq P(X<d) $ (when $i=1$) is as follows. 
\begin{itemize}
\item[(i)] in case $r_1 < d/s$, $r_i \in (0,d/s)$;   
\item[(ii)] in case $r_1 \geq d/s$, $r_i \in (r_m,r_M)$.
\end{itemize}
\end{proposition}

\subsection{The probability of joint default} 

To compute the probability of joint default using the geometry of Fig 2 in Wagner, we have to keep in mind that the two lines switch positions according to whether $r_1+r_2 < 1$ or $r_1+r_2 \geq 1.$ In each case one only has to compute the areas of a couple of triangles and add it to  the area of the square. The result is
\begin{equation}\label{eq3}
P\left\{\nu(1)\leq d, \nu(2)\leq d\right\}=\left\{\begin{array}{cc}
						\frac{1}{2}\Big(\frac{d}{s}\Big)^2\left[\frac{1}{1-r_1}+\frac{1}{1-r_2}\right]& r_1+r_2 <1 \\               
\frac{1}{2}\Big(\frac{d}{s}\Big)^2\left[\frac{1}{r_1}+\frac{1}{r_2}\right]  & 1 \leq r_1+r_2\end{array}\right.
\end{equation}

Note that the first identity increases as any (or both) $r_1,r_2$ increase, and that the second increases when any (or both) $r_1,r_2$ decrease. The boundary between the two domains in (\ref{eq3}) is the line $r_1+r_2=1,$ or equivalently $r_1=1-r_2.$ If we put $r_1=1-r_2=r$ the two alternatives in (\ref{eq3}) become equal, that is 
$$P\left\{\nu(1)\leq d, \nu(2)\leq d\right\}=\frac{1}{2}\Big(\frac{d}{s}\Big)^2\left[\frac{1}{1-r}+\frac{1}{r}\right].$$
To finish, note that $r=1/2$ minimizes the right hand side, that is the probability of joint default occurs when the two banks hold exactly the same position and it is $2d^2/s^2.$ As it is was noted by Wagner,  
$$P\left\{\nu(1)\leq d, \nu(2)\leq d\right\} > P(X \leq d, Y \leq d)=\frac{d^2}{s^2}$$

Hence, banks focusing only on the level of systemic risk, will choose not to diversify.

\subsection{The correlation between $\nu_1$ and $\nu_2$}
It is not hard to see that, since $X$ and $Y$ are identically distributed, the covariance between $\nu_1$ and $\nu_2$ is given by
$$Cov(\nu_1,\nu_2) = (r_1 + r_2 - 2r_1r_2)\sigma^2,\,\,\mbox{where}\;\;\sigma^2=\sigma^2(X)=\sigma^2(Y).$$
We now state:
\begin{theorem}\label{cov}
Consider the function $w:\bbR^2 \to \bbR$ define by $w(r_1,r_2)=r_1+r_2-2r_1r_2.$ then\\
{\bf 1)} The Hessian matrix $\partial^2 w/\partial r_i\partial r_j$ has eigenvalues $\pm 2.$\\
{\bf 2)} The function $w$ is analytic on the whole plane, i.e., $\partial^2 w/\partial^2r_1 +\partial^2 w/\partial^2 r_2 = 0.$\\
It is continuous in $[0,1]^2,$ then by the maximum principle, it extremal values are achieved on the boundary $\partial[0,1]^2$ of the unit square. 
\end{theorem}
The first item explains why $r_1=r_2=1/2$ is a saddle point. The first derivatives vanish there, by (2) it is not an extremal point. 
As a matter of fact $cov(\nu_1,\nu_2)=0$ at $r_1=r_2=0$ and at $r_1=r_2=1,$ in which cases the investors (or banks) are statistically independent. Similarly, $cov(\nu_1,\nu_2)=\sigma^2$ at $r_1=1, r_2=0$ or at $r_1=0,r_2=1,$ and in both of these cases the investors (banks) are again independent. In all other cases they correlated. This further explains why the probability of joint default of the diversified investors is larger that of individual, statistically independent investors. \\

\par\vspace{6pt}

To emphasize, when $r_1=r_2=1$ and when $r_1=r_2=0$, we get that $cov(\nu_1,\nu_2)=0.$ This makes sense if one looks at the return of the entire banking system. Since in these two cases the expected returns of the banking system is just $x+y$, then the covariance of these two i.i.d's must be zero. On the other hand, when $r_1=1, r_2=0$ or at $r_1=0,r_2=1$; the covariance is $cov(\nu_1,\nu_2)=\sigma^2$. From the point of view of the entire banking system, the expected return is either $2E[X]$ or $2E[Y]$. Since the covariance of any distribution with itself is the variance, we get in this case that $cov(\nu_1,\nu_2)=\sigma^2$. 

\par\vspace{6pt}

 Observe as well, that when $r_1=r_2=1$ then $r_1+r_2>1$. Hence, the probability of failure in this case is given by, 
\[\frac{1}{2}\Bigg(\frac{d}{s}\Bigg)^2\Bigg[\frac{1}{r_2}+\frac{1}{r_1}\Bigg]\]
If we ask when do we get a smaller probability of systemic failure in this case, one can see that this happens when $r_1=r_2=1$ (no diversification). If $r_1=r_2=0$ (i.e., $r_1+r_2<1$), we then have that the probability of the system failing is, 
\[\frac{1}{2}\Bigg(\frac{d}{s}\Bigg)^2\Bigg[\frac{1}{(1-r_2)}+\frac{1}{(1-r_1)}\Bigg]\] 
If we ask again when do we get a smaller probability of systemic failure in this case, one can see that this occurs when $r_1=r_2=0$ (no diversification). 

\par\vspace{6pt}

To finish, observe that within the framework of the model, if for example $r_1=1$ and $0<r_2<1,$ then $0<1+r_2 - 2r_2=1-r_2<1,$ that is, the covariance between $\nu_1$ and $\nu_2$ is never negative. Therefore: 
\begin{equation}\label{eq4}
Var(\nu_1+\nu_2) = Var(\nu_1) + Var(\nu_2) + Cov(\nu_1,\nu_2) \geq Var(\nu_1) + Var(\nu_2).
\end{equation}
Using the variance, or standard deviation, as a proxy for risk in this case would indicate that diversification is {\textquotedblleft{at least as risky}"} as the no diversification case. However, it must be bear in mind that the standard deviation is not a good choice for a risk measure because it penalizes - symmetrically - both positive and negative deviations from the mean. What is interesting about (\ref{eq4}) is that by taking the covariance as a way to assess the similarity between the bank's portfolios, we obtain a complementary way for better understanding the dark side of diversification in Wagner's (2010) model. The more similar the banks are, the higher the covariance and the higher the risk of joint failure. 

\section{Risk measurement analysis}
From the point of view of risk analysis, the essential issue is how much one looses with a given probability. We shall carry out a detailed analysis for any given bank with respect to the value at risk (VaR), which is the most widely used risk measure by financial institutions and regulatory agencies. After that we shall address the case of systemic risk \footnote{Let us mention that if we consider a coherent or convex risk measure $\rho$ (see the appendix for a formal definition), like the expected shortfall, then it is necessarily true that the risk of the diversified portfolio is smaller that of any of its components.}. 

\par\vspace{6pt}

Recall that VaR provides us with the smallest loss with a given confidence level, or the smallest loss with a given probability if you prefer. The VaR was approved in the  mid 1990's by regulators as a valid approach for calculating capital reserves needed to cover market risk. The Basel Committee on Banking Supervision has released several amendments, but the capital reserves that financial institutions are required to keep could be based on VaR numbers computed by an in-house risk management system. The use of VaR for the purpose of capital requirements has been discussed in the literature in relation to business cycles and heavy-tailed distributions (see Adrian and Shin (2013), Rossingolo, Fethi and Shaban (2012), and P\'erignon, Deng, and Wang (2008)). Our use of VaR is applied specifically in the context of Wagner's model (2010) where there are no considerations of future periods and, therefore, there is no room for business cycles. 

\par\vspace{6pt}

\subsection{The one bank case}
We shall see that with respect to the VaR for appropriate values of $r,$ for any bank the diversified portfolio given by $\nu=(1-r)X+rY$ has a smaller risk measure, thus diversification pays off. We should stress that even though the value at risk is not a coherent risk measure, it is still enforced by regulators. In our case it happens to decrease with diversification.

\par\vspace{6pt}

But before that, as the variables $X$ and $Y$ considered above are positive, in order to talk about losses, we will redefine them so that they are negative below the default threshold:
$$X \longrightarrow X_1 = X-d.$$
$$Y \longrightarrow Y_1 = Y-d.$$
So, an investor holding $X_1$ experiences losses when $X<d.$ Similarly, we shall replace $\nu$ by 
$$\nu_1(r) \longrightarrow R(r) = (1-r)X_1 + rY_1$$
\noindent and then $R_1<0\Leftrightarrow \nu_1(r)<d.$ Given a confidence level $0<\alpha<1)$ (or probability of loss $1-\alpha$,) in our current model the value at risk at level $\alpha$ of $X_1$ is defined by
\begin{equation}\label{var1}
P(X_1 \leq -VaR_\alpha(X_1)) = 1-\alpha.
\end{equation}
That is, it is the $(1-\alpha)-$quantile if $X_1.$ The value at risk satisfies the following two properties (see \cite{FS}). Let $W$ be a random variable modeling a risky financial position, then :\\
{\bf 1)} If $m$ in any real number, then $VaR_\alpha(W+m)=VaR_\alpha(W)-m.$\\  
{\bf 2)} For any $\lambda\geq 0$ we have $VaR_\alpha(\lambda W)=\lambda VaR_\alpha(W).$

Using these two properties, since $X_1=X-d$ and $X\sim U[0,s]$ we have 
$$\frac{1}{s}[d-VaR_\alpha(X_1)] = 1-\alpha\;\;\;\mbox{and} \;\;\; -VaR_\alpha(X) = s(1-\alpha).$$
We collect these remarks under
\begin{lemma}\label{lemf}
$$P\big(X_1 \leq -VaR_\alpha(X_1)\big) = 1-\alpha \;\;\Rightarrow\;\; VaR_\alpha(X_1) = d-s(1-\alpha).$$
\end{lemma}
Similarly, since $R_1(r)=\nu_1(r)-d,$ it is clear that the set $\{R_1(r) \leq -VaR_\alpha(R_1)\}= \{\nu_1(r) \leq d - VaR_\alpha(R_1)\}$ is a triangle with base $(d-VaR_\alpha(R_1))/(1-r)$ and height $(d-VaR_\alpha(R_1))/r,$ but as in Section 2, the problem is that one of these two points might be larger than $s.$ To study this case, first note that from the definition of $X_1,Y_1$ that  $-d \leq R(r) \leq s-d$ with probability 1. Since we need to find $V$ such that $P(R(r) \leq -V) = 1-\alpha$ the former implies that $-d\leq-V,$ or there is no loss than the worst possible loss. 

\par\vspace{6pt}

To describe the region $\{\nu_1(r) \leq d-V\}$ let us first use $K=d-V$ for short and consider the line $(x+y)/2=K.$  This line intersects the square at the points $(2K,0)$ and $(0,2K)$ and as since we suppose that $2d<s,$ then $2K=2(d-V)<s.$ Note now that the lines $(1-r)x+ry=K$ are pivoted at $(K,K)$ and rotate counterclockwise if $r$ increases beyond $1/2$ or clockwise if $r$ decreases below $1/2.$ The problem is that to know where the line intersects the axes, we would have to know $V=d-K$ which depends on $\alpha$ and $r$ to begin with. Anyway, supposing $K$ were known, the relationship between $V,r,d,s$ and $\alpha$ is easy to work out, and summed up (recall that $V=VaR_\alpha(R_1)$) in the following result.

\begin{lemma}\label{lems}
If as above, $V=VaR_\alpha(R_1),$ then:
\begin{equation}\label{rel2}
V =\left\{\begin{array}{cc}
			d-s[1-r\alpha] &\mbox{when}\;\; s\leq (d-V)/(1-r)\;\\
			d-\big(2(r(1-r)(1-\alpha)\big)^{1/2}&\mbox{when}\;\;(d-V)/(1-r)\leq s\;\& \;\leq (d-V)/r\leq s\\
                         d-s[1-(1-r)\alpha] &\mbox{when}\;\; s\leq (d-V)/r.\end{array}\right.  
\end{equation}
\end{lemma}

And to conclude, to describe the effect of diversification, we have:
\begin{theorem}\label{main}
With the notations introduced above, in the first and third cases of Lemma \ref{lems}, taking Lemma \ref{lemf} into account we have:
$$VaR_\alpha(R_1) < VaR_\alpha(X_1).$$
In the second case of Lemma \ref{lems}, the comparison with the result given in Lemma \ref{lemf} is split into two cases:

\begin{tabular}{ccc}
$VaR_\alpha(R_1) < VaR_\alpha(X_1)$ & according to  & $\sqrt{2(r(1-r)} > \sqrt{(1-\alpha)}$\\
$VaR_\alpha(R_1) > VaR_\alpha(X_1)$ & according to  & $\sqrt{2(r(1-r)} < \sqrt{(1-\alpha)}$
\end{tabular}
\end{theorem}

To explain why the second case is the interesting one, recall from Section 2 that in a neighborhood of $r=1/2,$ the probability of default of a diversified bank is less than its probability of default in the absence of diversification. Note as well that for $d-V>0$ the line $(1-r)x+ry=d-V$ lies to the left of the line $(1-r)x+ry=d.$ This line intersects the axes inside the $[0,s]^2$ square as long as $2d/s<\leq r \leq 1.$ And within that range the second condition in Lemma \ref{lems} holds. 

\par\vspace{6pt}

To finish, if we consider VaR with a high confidence level, $1-\alpha$ is a small number then for a large range of $r$ we have that $VaR_\alpha(R_1)< VaR_\alpha(X_1).$ 

\par\vspace{6pt}

Had we considered any  coherent or convex risk measure $\rho,$ diversification always reduces risk\footnote{See the appendix}, because then, regardless of the value of $r$ we have: 
$$\rho(R_1(r)) = \rho\big((1-r)X_1 + rY_1\big) \leq (1-r)\rho(X_1)+ r\rho(Y_1) = \rho(X_1)$$
because $\rho(X_1)=\rho(Y_1)$ since $X,Y$ are equally distributed. We emphasize that this a general result independent of the risk measure. If we needed to compute the risk explicitly, we would have to consider a specific risk measure. A standard example of a coherent risk measure is the expected short fall, defined by $\rho(R) =-E[R|R\leq VaR_\alpha(R)]$ for any continuously distributed risk $R,.$ In our case, the computation of this quantity is simple but lengthy.

\subsection{What about the systemic risk?}
There is an issue implicit is Wagner's work which is related to an open problem: How to measure risk to a random vector which describes a collective financial position. That is, how to measure risk of a collective financial position. In this case we have a random vector describing the collective individual risky positions. In Wagner's two bank system it would be $(R(r_1),R(r_2))$


If we proceed naively, and, as in the case of a financial entity that aggregates all its risks to obtain an aggregated risk, for the two bank system the result of the aggregation process lead us to:
$$R_{total} = R(r_1) + R(r_2) = (1-r_1+r_2)X_1 + (1-r_2+r_1)Y_1.$$
Observe that this corresponds to what Wagner calls ``the merger'' of the two banks in Section 5 of his paper. Again, note that if $\rho$ is any coherent or convex risk measure, without computing anything we can assert that
$$\rho(R_{total}) = \rho\Big(R(r_1) + R(r_2)\Big) \leq \rho(X_1) + \rho(Y_1).$$

Having settled for $R_{total}$ as characterizing the systemic risk, we can ask: What is $P(R_{total} \leq 0)$? Notice that
the definitions imply
$$P\big(R_{total} \leq 0\big) = P\big((1-r_1+r_2)X + (r_1+1-r_2)Y \leq 2d\big).$$
In Wagner's model, as long as $2d\leq s,$ this amounts to compute the area of a triangle with base $2d/s(1-r_1+r_2)$ and height $2d/s(1-r_2+r_1).$ The result is 
\begin{equation}\label{joint3}
P\big(R_{total} \leq 0\big) = \frac{2d^2}{s^2(1-r_1+r_2)(r_1+1-r_2)}.
\end{equation}
Note that the denominator has a maximum along the line $r_1=r_2,$ in which case 
$$P\big(R_{total} \leq 0\big) = \frac{2d^2}{s^2}.$$
If none of the two banks is diversified, the total position at risk is $X+Y-2d$ and a similar argument (or particularize at $r_1=r_2=0$ if you prefer) yields:
$$P\big(X+Y-2d \leq 0\big) = P\big(X+Y \leq 2d\big) = \frac{2d^2}{s^2}.$$
And we can gather the comments under
\begin{theorem}\label{last}
With the notations introduced above
$$ P\big(R_{total} \leq 0\big) \geq P\big(X+Y-2d \leq 0\big)$$
and the equality holds whenever $r_1=r_2,$ that is whenever the two banks chose to invest the same amounts in each other.
\end{theorem}
That is, if we look at the aggregate financial position of the system, the probability of loss before and after diversification is the same as long as the two banks use the same diversification. Or to use Wagner's phrasing: If the two banks use the same diversification, to diversify does not increase the probability of default.

\par\vspace{6pt}

To finish we have:
\begin{theorem}\label{last2}
Suppose that the two banks invest similarly in each other, that is $r_1=r_2,$ and $r$ satisfies the center condition in (\ref{rel2}) of Lemma \ref{lems} holds,  then
$$VaR_\alpha\big(R(r_1)+R(r_2)\big) \leq VaR_\alpha\big(R(r_1)\big) + VaR_\alpha\big(R(r_2)\big).$$
\end{theorem}
If we want to solve for $V,$ under our assumptions we have 
$$P\big(R(r_1)+R(r_2)\leq -V\big)=1-\alpha \;\;\Leftrightarrow \;\; P\big(X + Y \leq 2d-V\big)=1-\alpha.$$
Therefore 
$$\frac{(2d-Var_\alpha\big(R(r_1)+R(r_2)\big))^2}{2s^2} = 1-\alpha \;\;\;\Leftrightarrow\;\;\; Var_\alpha\big(R(r_1)+R(r_2)\big) = 2d- s\big(2(1 - \alpha)\big)^{1/2}.$$
Observe now that if the two banks invest $r_1=r_2=r$ in each other, and when $r$ satisfies the required conditions, and taking into account that $4r(1-r)\leq 1,$ a simple computation leads us to:
$$VaR_\alpha\big(R(r)\big) + VaR_\alpha\big(R(r)\big) \geq 2d -s\big(2(1-\alpha)\big)^{1/2} = VaR_\alpha\big(R(r_1)+R(r_2)\big).$$ Again, with the caveat that value at risk may not always decrease with diversification, in this case the conclusion that a regulator would appreciate is, let the banks invest the same amount in each other, then not only does the probability of systemic default not increase, but also the value at risk of the system is lower. 

\section{Discussion}
We have shown that, if one uses the probability of a systemic crisis as the basis for risk management then - as Wagner (2010) correctly pointed out - diversification comes at a cost (i.e., the dark side of diversification). If, instead, one uses VaR as a risk measure, then diversification for the banking system as a whole does not exhibit a dark side. In the first case, the analysis suggests a rationale for discouraging diversification. In the second case, as we have shown, the analysis suggests a rationale for encouraging it. Wagner (2010) argues that with respect to capital requirements, the first case would imply that banks with more diversified portfolios should be subject to higher capital charges. In the second case, it would imply that banks with more diversified portfolios are subject to lower capital charges. So, what metric should banks and regulators use? And why? 

\par\vspace{6pt}

If the rationale indicates that diversification ought to be discouraged by increasing capital requirements, then it would be sensible to ask how do banks respond to changes in capital requirements? And what are the costs and benefits for banks associated with a change in capital charges? For example, capital requirements may affect the liquidity and profitability of banks (see Blum (1999) and Tran, Lin, and Nguyen (2016)) and, as a consequence, affect negatively the probability of bank failure. Hence, when considering the increase in the likelihood of a systemic crisis due to diversification on one hand, and the increase in the probability of systemic crisis and higher capital requirements on the other; it can be argued that the rationale for discouraging diversification is less straightforward than what Wagner (2010) appeared to suggest. At the same time, the use of a popular measure like VaR in the context of Wagner's model does not impose a tradeoff between capital charges and the risk of the banking system as a whole. Therefore, after considering these elements in the context of a rationale for discouraging diversification, the dark side may not be so {\textquotedblleft{dark}"} after all.  

\par\vspace{6pt}

Our work can be extended in various directions. One line of further inquiry, is to link the systemic crisis in relation to {\textquotedblleft{diversity}"}\footnote{By diversity we mean portfolios that are different in the sense that they have a low or even negative correlation} rather than diversification; and to investigate the extent in which different measures of risk may lead -in the case of diversity - to less ambiguous decisions.




\section{Appendix: Short review of risk measures}
As we saw above, we are modeling risks by bounded random variables. Denote the underlying sample space by $(\Omega, \cF,P)$ and denote by $\cL$ the class of bounded random variables. Following F\"olmer and Schied's \cite{FS}, we state

\begin{definition}\label{rm1}
A function $\rho:\cL \to \bbR$ is said to be a monetary risk measure if it satisfies the following axioms:
\begin{eqnarray}
{\rm 1)}\;\mbox{Monotonicity: For}\;\;X_1,\,X_2\in\cL\;\;X_1\leq X_2 \Rightarrow \rho(X_2) \leq \rho(X_1).\\ \label{rm1.1}
{\rm 2)}\;\mbox{Cash Invariance: For}\;\;X_1\in \cL\;\;a \in \bbR\;\; \Rightarrow \rho(X + a) = \rho(X)-a.\\ \label{rm1.2}
\end{eqnarray}
\end{definition}
Requiring some extra properties one has
\begin{definition}\label{rm2}
A monetary risk measure $\rho$ is said to be convex if it satisfies
\begin{equation}\label{rm2.1}
\mbox{Convexity}\;\mbox{For}\; 0\leq\lambda\leq 1\;\;X_1,\,X_2\in\cL\;\;\Rightarrow \rho(\lambda X_1 + (1-\lambda)X_2) \leq \lambda\rho(X_1) + (1-\lambda)\rho(X_2).
\end{equation}
And a convex risk measure is called coherent whenever
\begin{equation}\label{rm2.2}
\mbox{Positive homogeneity}\;\;\; \mbox{For}\;\; 0 \leq \lambda\;\;\;X \in\cL\;\;\Rightarrow \rho(\lambda X) = \lambda\rho(X).
\end{equation}
\end{definition}
The expected shortfall at confidence level $\alpha$ is a nice coherent risk measure based on VaR is defined as follows. Let $X$ be a continuous random variable with finite expectation modeling some risky financial position.  Then:
$$ES_\alpha[X] = -E[X | X < VaR_\alpha(X)] = -\frac{1}{1-\alpha}\int_{-\infty}^{VaR_\alpha(X)}xdF_X(x).$$ 
 \noindent which is to be interpreted as the expected loss given that losses larger than the value at risk occur.

Comments: The value at risk is not coherent for it is non-necessarily monotone. Similarly, the standard deviation (or volatility, as usually called in finance) is a positively homogeneous convex function, but it is not monotone and cash invariant. To finish we mention that in our models, the computation of the expected shortfall involves computing integrals of the type
$$\int_{\{ax+by\leq K\}}xdxdy.$$
But we do not need the exact result of such computations. We only need to know that $ES_\alpha[\nu_1+\nu_2] \leq ES_\alpha[\nu_1]+ ES_\alpha[\nu_2]$ because that is a generic property of the expected shortfall.

 \end{document}